\documentclass[5p,twocolumn,superscriptaddress,nofootinbib,preprintnumbers]{elsarticle}

\usepackage{amsmath,amssymb,amsfonts}
\usepackage{geometry}
\usepackage{graphicx}
\usepackage[dvipsnames]{xcolor}
\usepackage{bm,tensor,physics}
\usepackage{hyperref}
\usepackage{etoolbox}
\apptocmd{\thebibliography}{%
  \setlength{\itemsep}{0pt}%
  \setlength{\parsep}{0pt}%
}{}{}
\hypersetup{colorlinks=true,urlcolor=blue,citecolor=blue,linkcolor=blue}
\newcommand{\be}{\begin{equation}}
\newcommand{\ee}{\end{equation}}
\newcommand{\bea}{\begin{eqnarray}}
\newcommand{\eea}{\end{eqnarray}}

\newcommand{\lp}{\left (}
\newcommand{\rp}{\right )}

\journal{Physics of the Dark Universe}
\biboptions{sort&compress}
\begin{document}
\makeatletter
\def\ps@pprintTitle{%
  \let\@oddhead\@empty
  \let\@evenhead\@empty
  \let\@oddfoot\@empty
  \let\@evenfoot\@oddfoot
}
\makeatother
\begin{frontmatter}

\title{The unavoidable de Sitter fate of a scale-invariant Universe}

\author[trento,tifpa,iem]{Chiara~Cecchini\corref{cor1}}
\ead{chiara.cecchini@unitn.it}

\author[trento,tifpa]{Massimiliano~Rinaldi}
\ead{massimiliano.rinaldi@unitn.it}

\cortext[cor1]{Corresponding author.}

\address[trento]{Department of Physics, University of Trento, Via Sommarive 14, 38122 Povo (TN), Italy}

\address[tifpa]{Trento Institute for Fundamental Physics and Applications (TIFPA) -- INFN, Via Sommarive 14, 38122 Povo (TN), Italy}

\address[iem]{Instituto de Estructura de la Materia (IEM), CSIC, Serrano 121, 28006 Madrid, Spain}

\begin{abstract}
We consider a very general scale-invariant scalar-tensor theory of gravity and its flat cosmological solutions. We show that any stable configuration with non-degenerate gravitational dynamics carries a non-vanishing cosmological constant, unless the quartic self-coupling of the scalar field vanishes. 
Since this condition is not protected against radiative corrections, a residual cosmological constant is expected as a generic and robust prediction of this class of theories.
This result suggests that dark energy may be a natural consequence of an early scale-invariant phase of the Universe.
\end{abstract}


\end{frontmatter}

\emergencystretch=2em
\hyphenpenalty=1000

\section{Introduction}\label{sec:intro}

The observed vacuum energy density, $\rho_\Lambda \sim 10^{-47}\,\text{GeV}^4$, is suppressed by roughly 120 orders of magnitude relative to the naive estimate from Planck-scale zero-point energies~\cite{Zeldovich_1968,Weinberg:1988cp,Padmanabhan:2002ji}. 
The discovery of cosmic acceleration~\cite{SupernovaSearchTeam:1998fmf,SupernovaCosmologyProject:1998vns} confirmed that $\Lambda$ is nonzero and dominates the energy content of the Universe today, yet no fundamental theory comes close to explaining its observed value, making the cosmological constant problem one of the deepest open questions in theoretical physics~\cite{Carroll:2000fy,Peebles:2002gy,Weinberg:2000yb}.
The issue is further sharpened by radiative instability: quantum corrections reintroduce a large vacuum energy at every loop order, so that any cancellation imposed at tree level must be repeated order by order~\cite{Weinberg:1988cp,Martin:2012bt,Burgess:2013ara}. Before addressing the quantum aspect of the problem, one should first ask whether the classical theory itself predicts a vanishing vacuum energy in its ground state. If it does not -- if fine-tuning is already required at tree level -- then the problem is present before quantization and cannot be deferred to a quantum mechanism.

A natural strategy is to invoke a symmetry that forbids a cosmological constant at the classical level. Scale invariance -- the absence of any intrinsic dimensionful parameter in the fundamental action -- represents the most natural candidate, since a strictly scale-invariant theory admits no preferred energy scale at which to anchor the vacuum energy.
This idea has a long history, dating back to early works on induced gravity \cite{Zee:1978wi,Cooper:1981byv,Fujii:1982ms,Adler:1982ri}, in which the gravitational coupling is not a fundamental parameter, but rather arises dynamically through the vacuum expectation value of a scalar field.
In these models, all mass scales -- including the Planck mass -- are generated through spontaneous symmetry breaking rather than being introduced by hand. 

The rationale for scale invariance as a fundamental principle extends well beyond the dynamical generation of masses. In four spacetime dimensions, requiring the absence of dimensionful couplings restricts the Lagrangian to operators of mass dimension four, making scale invariance a highly predictive organizing principle. 
A concrete example is provided by the Starobinsky model~\cite{Starobinsky:1980te}, in which the leading modification of the Einstein-Hilbert Lagrangian is a quadratic $R^2$ term. This is the simplest higher-order curvature correction that is both ghost-free and one-loop renormalizable; notably, the $R^2$ term itself is exactly scale invariant, unlike the Einstein-Hilbert term it corrects.
The empirical success of Starobinsky inflation naturally suggests that scale invariance may be realized as an approximate symmetry of gravity in the high-curvature regime. 
In particle physics, the Standard Model is approximately scale invariant above the electroweak scale, and the origin of the electroweak scale itself may be tied to the dynamical breaking of this symmetry \cite{Bardeen:1995kv,Shaposhnikov:2008xi,Salvio:2014soa}.
At a more fundamental level, Wetterich has proposed \cite{Wetterich:2019qzx} that scale invariance may provide a unifying effective field theory framework connecting early-Universe ultraviolet (UV) physics, the Standard Model, and a late-time infrared (IR) regime, with all physical scales emerging along a single renormalization-group flow.
These considerations provide strong motivation for the systematic implementation of scale invariance in gravitational and cosmological settings; see, e.g., ~\cite{Cooper:1981byv,Finelli:2007wb,Garcia-Bellido:2011kqb,Salvio:2014soa,Rinaldi:2015uvu,Ferreira:2018qss,Cecchini:2024xoq,Kannike:2015apa}.

Despite these attractive features, scale-invariant models generally predict a non-vanishing residual cosmological constant after symmetry breaking, even though no explicit mass scale appears in the classical action. 
In full analogy with the gauge hierarchy problem, where the large separation between the electroweak and Planck scales is translated into a hierarchy among dimensionless couplings, the cosmological constant problem in scale-invariant theories is not eliminated but effectively shifted to a fine-tuning of the dimensionless parameters of the theory ~\cite{Garcia-Bellido:2011kqb, Shaposhnikov:2008xi}. 

Several routes have been explored to address this issue within or beyond the scale-invariant framework. 
One possibility, explored in the context of cosmon inflation~\cite{Wetterich:2013wza}, is that quantum corrections introduce an intrinsic renormalization scale $\bar\mu$ that explicitly breaks scale invariance. The cosmological constant is then driven to zero dynamically as the scalar field evolves toward an IR fixed point where scale symmetry is restored. 
A second direction is to enlarge the symmetry group to local conformal invariance, since the Ward identities of any spontaneously broken conformally invariant theory ensure a vanishing cosmological constant to all orders in perturbation theory, see~\cite{Antoniadis:1984kd,Cadoni:2006ww}.
Another possibility is to extend the field content of the theory by introducing additional scalar degrees of freedom, as explored, e.g., in~\cite{vandeBruck:2021xkm}. 

More recently, the pathological Minkowski limit of pure $R^2$ gravity has been revisited using a thermal analogy, providing a new perspective on the vacuum structure of scale-invariant gravity~\cite{Faraoni:2026wlc}.

In this work, we ask whether a vanishing cosmological constant can be achieved without enlarging the field content or symmetry of the theory. We consider the most general scale-invariant scalar-tensor action of the form $J(R/\phi^2)$, where $R$ is the Ricci scalar and $\phi$ a real scalar field.
We impose no restrictions on the non-minimal coupling, scalar potential, or kinetic term beyond analyticity of $J$ and scale invariance.

We show that every stable fixed-point configuration with non-degenerate gravitational dynamics necessarily carries a non-vanishing cosmological constant, unless the quartic self-coupling of the scalar field vanishes identically. This result is not a shortcoming of specific models, but rather a model-independent feature of scale-invariant gravity.

\section{A no-go result for the residual cosmological constant}\label{sec:nogo}

We consider the following action for a single-field scalar-tensor theory, which is also scale-invariant:
\be
\label{eq:JFaction}
S_{\rm{JF}} = \int \dd^4 x\sqrt{-g}\left[J\left(\dfrac{R}{\phi^2}\right)\phi^4 - \dfrac{1}{2} g^{\mu\nu}\partial_\mu \phi\partial_\nu\phi\right], 
\ee
where the subscript JF denotes the Jordan frame and $J$ is a generic function of the dimensionless argument $R/\phi^2$. 
As will be clarified in the following, the choice of a canonical kinetic term for the scalar field is made without loss of generality. 

The form of the function $J$ is arbitrary. We only require $J$ to be analytic in its argument for any non-vanishing value of $\phi$.

The trace of the Einstein equations reads
\be\label{eq:trace}
3\Box (\phi^2 J') = \phi^4(2J-XJ')-\dfrac{1}{2}(\partial \phi)^2, 
\ee
with a prime denoting the derivative of $J$ with respect to its argument $X\equiv R/\phi^2$.
The Klein-Gordon equation for $\phi$ is 
\be\label{eq:KG}
\Box\phi + 2\phi^3(2 J-XJ') = 0\,.
\ee
To demonstrate our claim, it is convenient to transform the action \eqref{eq:JFaction} to the Einstein frame. 
Since the Jordan frame gravitational coupling is field-dependent, the identification of a residual cosmological constant is most transparent in the Einstein frame, where gravity is minimally coupled and the vacuum energy is unambiguously given by the value of the effective potential at the fixed point.

To carry out this transformation, we first rewrite the action \eqref{eq:JFaction} in an equivalent form by introducing an auxiliary Lagrange multiplier field $\chi$ and an auxiliary scalar field $\psi$, both of mass dimension two, generalizing the procedure of \cite{Ferreira:2018qss} (see also \cite{Garcia-Bellido:2011kqb} for an earlier application in the context of Higgs-dilaton cosmology):
\be
\label{eq:S_J}
    S_{\rm{JF}} = \int \!\dd^4 x \sqrt{-g}\biggl[J\!\biggl(\frac{\psi}{\phi^2}\biggr)\phi^4+\chi(R-\psi)-\frac{1}{2} g^{\mu\nu}\partial_\mu \phi\partial_\nu\phi\biggr].
\ee
Varying the action \eqref{eq:S_J} with respect to $\chi$ enforces the constraint $R=\psi$, showing that Eq.~\eqref{eq:S_J} is on-shell equivalent to Eq.~\eqref{eq:JFaction}. Similarly, variation with respect to $\psi$ gives
\be
\label{eq:KGpsi}
J'\left(\dfrac{\psi}{\phi^2}\right) = \dfrac{\chi}{\phi^2}\,.
\ee
where the prime now denotes the derivative with respect to the argument $X=\psi/\phi^2$.

The auxiliary formulation \eqref{eq:S_J} also makes the scale symmetry of the theory more transparent, allowing for a compact derivation of the associated Noether current. 
The resulting current is~\footnote{We consider an infinitesimal scale transformation: $\phi \to e^{-\epsilon}\phi$, $g_{\mu\nu}\to e^{2\epsilon}g_{\mu\nu}$, $\psi \to e^{-2\epsilon}\psi$, $\chi \to e^{-2\epsilon}\chi$, where $\epsilon$ is an infinitesimal constant. The associated Noether current is obtained following the standard procedure \cite{Coleman:1985rnk}; see also \cite{Garcia-Bellido:2011kqb,Ferreira:2018qss} for its application in scale-invariant gravity.}
\be
\label{eq:current}
K^{\mu} = \partial^{\mu} K, \qquad 
K = \dfrac{\phi^2}{2} + 6\chi = \dfrac{\phi^2}{2}\lp 1+2J'(X) \rp,
\ee
where the last equality holds on shell, using Eq.~\eqref{eq:KGpsi}. The equations of motion guarantee that the current is covariantly conserved, i.e., $\nabla_{\mu}K^{\mu}=0$. In a spatially flat Friedmann-Lemaître-Robertson-Walker (FLRW) spacetime, this condition implies
\be
K(t) = c_1 + c_2 \int_{t_0}^{t} \dfrac{\dd t'}{a^3(t')},
\ee
where $a(t)$ is the scale factor and $c_i$ are integration constants. In an expanding universe, the integral term decays and $K(t)$ approaches a constant value, thereby signaling the breaking of scale symmetry.

Eq.~\eqref{eq:current} admits a transparent geometric interpretation.
The on-shell constraint $K(t)\to c_1$ defines, in the field space
spanned by $(\phi, \sqrt{\chi})$, an elliptic region
\be
\label{eq:ellipse}
\dfrac{\phi^2}{2} + 6\chi = c_1\,,
\ee
where we restrict ourselves to $\chi > 0$.
At late times, the dynamical trajectory in the field space is, therefore, confined to this bounded region. The ellipse degenerates to the single point $(\phi, \sqrt\chi) = (0,0)$ -- corresponding to the unbroken, symmetric phase -- if and only if $c_1 = 0$.
For any set of initial conditions satisfying $c_1 \neq 0$, the late-time dynamics is confined to a non-degenerate ellipse in the $(\phi, \sqrt{\chi})$ plane, and the fixed point necessarily lies at some  $\phi_* \neq 0$.

The absence of viable stable configurations with $\phi_* = 0$ is further supported by the dynamical mass generation mechanism typical of scale-invariant models \cite{Cooper:1981byv, Fujii:1982ms}. When spontaneous symmetry breaking occurs, the system evolves toward a stable fixed point at which an effective mass scale emerges. In the present setup, as in more minimal scale-invariant models \cite{Cooper:1981byv, Rinaldi:2015uvu, Ferreira:2018qss}, this scale is set by the vacuum expectation value of the scalar field through the combination $\phi^2 J'(X)$, which determines the effective gravitational coupling and corresponds, on shell, to the coefficient of the term $R$ in the Jordan-frame action~\eqref{eq:S_J}. A fixed point with $\phi_* = 0$ would imply the absence of any dynamically generated scale.
This is not merely a formal shortcoming: at $\phi_* = 0$ the effective Planck mass $M_{pl}^2~=~2J'(X)\phi^2$ vanishes, and with it the kinetic term for tensor perturbations. The graviton propagator is ill-defined, and the configuration lies outside the domain of validity of the theory as a classical field theory.

For the above reasons, from now on we will consider fixed-point configurations with $\phi_* \neq 0$. 

By performing a conformal transformation on the metric $\tilde{g}_{\mu\nu} = \Omega^2g_{\mu\nu}$, and further choosing $\Omega^2 = 2\chi/M^2$, we finally obtain the Einstein-frame action
\be
\begin{split}
S_{\rm{EF}} = \int \dd^4 x \sqrt{-\tilde{g}}\biggl[ &\dfrac{M^2}{2}\tilde{R} - \dfrac{3M^2}{4\chi^2}\tilde{g}^{\mu\nu}\partial_{\mu}\chi\partial_{\nu}\chi +\\
&\, -\dfrac{M^2}{4\chi^2}\tilde{g}^{\mu\nu}\partial_{\mu}\phi\partial_{\nu}\phi - V_{\rm{EF}}\biggr], 
\end{split}
\ee
where a tilde denotes quantities in the Einstein frame (EF), and the effective potential can be written in the compact form
\be
\label{eq:VEF}
V_{\rm{EF}} = \dfrac{M^4}{4}\dfrac{J'(X)X-J}{J'(X)^2}. 
\ee
Note that the mass parameter $M$, introduced for dimensional consistency, is redundant and preserves the scale invariance of the theory \cite{Weinberg:1995mt, Rinaldi:2015uvu}.
Moreover, to preserve signature, causal structure, and invertibility of the metric, the conformal transformation must satisfy $\Omega^2>0$, which is consistent with the fact that we restrict ourselves to $\chi>0$. 
Employing Eq.~\eqref{eq:KGpsi}, this condition reduces on shell to
\be
\label{eq:conformal_factor}
\Omega^2 = \dfrac{2 \phi^2}{M^2} J'(X) >0. 
\ee
In particular, if $J'(X)$ or $\phi$ vanishes at some point, the metric in the Einstein frame becomes degenerate, and the Einstein frame ceases to be a suitable description for the model at that point. 

From Eqs.~\eqref{eq:trace} and \eqref{eq:KG}, it follows that any fixed point at $X=X_*$ satisfies
\be\label{eq:fpcond}
J'(X_*) X_* = 2 J(X_*)\,, 
\ee 
for $\phi_*\neq0$. 
We note that this relation remains unchanged if the canonical kinetic term in the action~\eqref{eq:JFaction} is replaced by a general $K(X)(\partial\phi)^2$, with $K$ an arbitrary function of $X$, since the fixed-point condition is evaluated at vanishing time derivatives of the fields.
 
To investigate the conditions under which a residual cosmological constant is absent, we examine when the Einstein-frame potential vanishes at the fixed point. Substituting the fixed-point condition~\eqref{eq:fpcond} into the expression for the potential~\eqref{eq:VEF}, one obtains
\be
\label{eq:VEF_fp}
V_{\rm{EF}}\big|_{X_*} = \dfrac{M^4}{4}\dfrac{J(X_*)}{J'(X_*)^2}\,,
\ee
which, for $J'(X_*) \neq 0$,~\footnote{If $J'(X_*) = 0$, the conformal factor in Eq.~\eqref{eq:conformal_factor} vanishes and the transformation to the Einstein frame is singular. Analogously, the effective Planck mass $M_{pl}^2 = 2J'(X_*)\phi_*^2$ vanishes in the Jordan frame. This case is discarded on the same grounds discussed above.} vanishes if and only if $J(X_*) = 0$. 
From Eq.~\eqref{eq:fpcond}, this condition is realized for $X_* = 0$. Since we assumed $J$ to be analytic, we can expand it around $X = 0$ as 
$J(X) = J(0) + J'(0)X + \ldots$, so that the Lagrangian density reads
\be
\label{eq:Jexpansion}
J(R/\phi^2)\phi^4 = J(0)\,\phi^4 + J'(0)\,R\phi^2 + \ldots\,,
\ee
where we identify $J(0)$ with the quartic self-coupling $\lambda/4$ and $J'(0)$ with the non-minimal coupling to gravity.
A vanishing cosmological constant in the Einstein frame therefore requires $J(0) = 0$, i.e., $\lambda = 0$, while maintaining 
$J'(0) \neq 0$ for a non-degenerate gravitational sector. 

We have therefore shown that, for theories of the form \eqref{eq:JFaction}, a vanishing cosmological constant at the fixed point requires $\lambda = 0$ -- a condition on the coupling constant itself, which cannot be realized through the classical evolution of the fields.
This requirement can be interpreted in two ways. 

The first is to set the quartic self-interaction to zero directly at the level of the classical Lagrangian in Eq.~\eqref{eq:JFaction}. 
However, this choice is not stable once quantum corrections are taken into account. Even if the quartic coupling is absent at tree level, radiative corrections generically generate a $\phi^4$ operator, so that $\lambda$ is induced through loop corrections unless protected by an additional symmetry~\cite{Vicentini:2019etr}. In this sense, imposing $\lambda=0$ at the classical level does not provide a technically natural realization of the condition required to eliminate the residual cosmological constant.

An alternative possibility is that the quartic coupling runs with the renormalization scale, $\lambda=\lambda(\mu)$, and vanishes dynamically at some scale $\mu_*$. In the present setup, the natural scale associated with the fixed point is set by the dynamically generated gravitational scale, which is expected to be close to the Planck mass. Therefore, one could imagine a scenario in which $\lambda(\mu_*)=0$ for $\mu_* \sim M_{pl}$. In this case, the fixed-point condition $\lambda=0$ would be satisfied only at that specific scale. The same conclusion was derived in \cite{Salvio:2014soa}. A situation of this kind is reminiscent of the behavior of the Higgs quartic coupling in the Standard Model. Renormalization-group analysis show that the Higgs self-coupling $\lambda_H(\mu)$ decreases with energy and approaches zero near the Planck scale for the measured values of the Higgs and top quark masses~\cite{Elias-Miro:2011sqh, Degrassi:2012ry, Buttazzo:2013uya}. Beyond this scale $\lambda_H$ turns negative, destabilizing the Higgs potential at large field values; during inflation, quantum fluctuations may drive the field beyond the potential barrier, rendering the electroweak vacuum metastable~\cite{Ellis:2009tp, Franciolini:2026jat}. 
In analogy to this situation, one could imagine that the quartic coupling in the present model follows a renormalization-group trajectory such that $\lambda(\mu)$ crosses zero near the dynamically generated gravitational scale. However, as in the Standard Model case, realizing this behavior requires a strict balance among the microscopic parameters controlling the running, and therefore does not eliminate the underlying fine-tuning problem.

Since neither interpretation provides a satisfactory resolution within the single-field theory, a vanishing cosmological constant can only be achieved by extending the field content of the theory. One such possibility was investigated in~\cite{vandeBruck:2021xkm}, where a second scalar field $\sigma$ is introduced in the Jordan frame, and the quartic self-interaction is replaced by
\be
\dfrac{\lambda}{4}\phi^4 \longrightarrow
\dfrac{\lambda}{4}\left(\phi^2 - \sigma^2\right)^2,
\ee
which vanishes for $\phi_*^2 = \sigma_*^2$ without requiring $\phi_* = 0$. In this case, the fixed-point condition for a vanishing cosmological constant is satisfied along a flat direction in the two-field space, rather than at a single point, and the no-go result derived above is evaded.

Finally, we note that our result establishes the presence of a non-vanishing residual cosmological constant, while leaving its magnitude entirely unconstrained. The value of $V_{\rm EF}$ at the fixed point, given by $M^4 J(X_*)/(4J'(X_*)^2)$, is completely determined by the dimensionless parameters entering the function $J$, and there is no mechanism within the classical theory to make it small. In this sense, scale invariance does not solve the cosmological constant problem, but rather reformulates it: the smallness of \(\Lambda\) is translated into a question of why certain dimensionless couplings, or ratios among them, take such small values.

\section{Conclusion}\label{sec:conclusion}

In this work, we have investigated the fixed-point structure of a general scale-invariant scalar-tensor theory of the form $J(R/\phi^2)$, assuming only analyticity of the function $J$. A central ingredient of the analysis is the conservation of the Noether current associated with scale invariance. This constraint restricts the late-time evolution to a bounded region of field space, excludes the symmetric phase $\phi_* = 0$ for generic initial conditions, and ultimately links the value of the cosmological constant to the dimensionless couplings of the theory. Within this setup, we have shown that any stable fixed point with non-degenerate gravitational dynamics necessarily corresponds to a de Sitter solution in the Einstein frame. The only exception arises when the quartic self-coupling of the scalar field vanishes, a condition that is not stable under radiative corrections.

Importantly, this conclusion does not rely on specific model choices, but applies to the full class of scale-invariant $f(R, \phi)$ theories, independently of the detailed form of the non-minimal coupling, the scalar potential, or the kinetic sector. At the classical level, it therefore identifies a non-vanishing vacuum energy as a structural feature of scale-invariant gravity. Within this perspective, the late-time accelerated expansion of the Universe emerges naturally, without the need to introduce an explicit cosmological constant term.

\section*{Acknowledgements}
CC and MR acknowledge support from the Istituto Nazionale di Fisica Nucleare (INFN) through the Commissione Scientifica Nazionale 4 (CSN4) Iniziativa Specifica ``Quantum Fields in Gravity, Cosmology and Black Holes'' (FLAG).  
CC acknowledges support from Fondazione Angelo Della Riccia. 

\section*{Data availability}

No data was used for the research described in this article.

\bibliographystyle{elsarticle-num}
\bibliography{biblio}

\end{document}